\documentclass[11pt]{article}
\usepackage[margin=1.1in]{geometry}
\usepackage{amsmath,amssymb}
\usepackage{xcolor}
\usepackage[colorlinks=true,linkcolor=black,citecolor=black,urlcolor=black]{hyperref}
\usepackage{cite}

\ifdefined\MARKED
  \newcommand{\new}[1]{{\color{blue}#1}}
\else
  \newcommand{\new}[1]{#1}
\fi

\renewcommand{\thesection}{\Roman{section}}

\makeatletter
\renewcommand{\p@subsection}{\thesection\,}
\makeatother

\newcommand{\E}{\mathcal{E}}
\newcommand{\U}{\mathcal{U}}
\newcommand{\dd}{\mathrm{d}}
\newcommand{\res}{{\rm res}}

\begin{document}

\begin{center}
{\Large\bf Energy of test objects on black hole spacetimes:\\[2pt] A brief review}\\[14pt]
{\large Alejandro Corichi}$^{1,2,*}$\\[8pt]
{\small\it
$^1$Centro de Ciencias Matem\'aticas, Universidad Nacional Aut\'onoma de M\'exico,\\
UNAM-Campus Morelia, A. Postal 61-3, Morelia, Michoac\'an 58090, Mexico\\[2pt]
$^2$Center for Fundamental Theory, Institute for Gravitation and the Cosmos,\\
Pennsylvania State University, University Park PA 16802, USA}
\end{center}
{\renewcommand{\thefootnote}{$*$}\footnotetext{Electronic address: corichi@matmor.unam.mx}}

\begin{abstract}
\noindent In this paper, we review the issue of defining energy for test particles on a background stationary
spacetime. We revisit the different notions of energy as defined by different observers. As is well
known, the existence of a time-like isometry allows for the notion of total conserved energy to be
well defined. We use this well known quantity to show that a gravitational potential energy can be
consistently defined. As examples, we study the case of the exterior regions of an asymptotically
flat black hole and of the $\Lambda>0$ Schwarzschild--de Sitter case, where an asymptotic region is not
available. We then consider the situation in which the test particle is absorbed by the black hole,
and analyze the energetics in detail\new{, making contact with the classic Geroch process}. In particular, we show that the notion of horizon energy as
defined by the isolated horizons formalism provides a satisfactory notion of energy compatible with
the particle's total conserved energy. With these choices, there is a global conservation of energy.
Finally, we comment on a recent proposal to define energy of the black hole as seen by a nearby
observer at rest, for which this feature is lost\new{, and show that the two descriptions differ by a Tolman redshift factor that leaves the change in horizon area, and hence the entropy, unchanged}.

\ifdefined\MARKED
\medskip

\noindent{\color{blue}\small [Revised version. Changes with respect to arXiv:1207.0887v3 are shown in blue.]}
\fi
\end{abstract}

\section{INTRODUCTION}

The concept of energy in general relativity is rather subtle. The Newtonian notion that
a quantity with the interpretation of energy must be conserved disappears once it
is realized that in general relativity there is no ``absolute'' background on which the theory
is defined. Instead, the geometry becomes a dynamical variable. As a consequence, the
notion of a gravitational energy density becomes ill defined. Still, there are solutions to
the equations of motion that are stationary, in the sense that there is a notion of time
invariance. It is then natural to expect that in those situations, there will be conserved
quantities associated to the motion of matter fields. That is indeed the case with the so-called
Komar quantities, constructed out of the Killing vector fields and the stress energy
tensor of matter. Down in the ladder of complexity, one might ask about the existence of a
conserved quantity for test particles moving on stationary space-times. Such a quantity has
been known for a long time, but its interpretation as a total energy of the particle is not
always found in the introductory treatment of the subject.

One of the purposes of this manuscript is to review and revisit this conserved quantity and
compare it to other notions of energy that have their origin in the special theory of relativity.
As we shall show, the conserved quantity can indeed be interpreted as a total energy, that
also contains information about the gravitational field. Thus, in certain circumstances, one
can indeed define the notion of a gravitational potential energy satisfying a set of desired
properties. The next question pertains to the case of the exterior of a black hole where the
spacetime is stationary and for which the horizon represents a boundary. We analyze in
detail the behavior of these quantities in the vicinity of the black hole. It is then natural
to consider the case when the particle falls into the \new{black} hole. Is there a notion of energy
associated to the black hole that can capture the intuitive idea that energy should not be
lost? Can one keep track of the energy of the in-falling object as being gained by the
black hole? As we shall see, the answer is in the affirmative. The natural notion of energy
within the isolated horizon formalism satisfies these properties. One can furthermore achieve
consistency with the energy balance, for both asymptotically flat and de Sitter black hole
spacetimes, provided we associate to the test particle the total (conserved) energy.

Still one might want to consider different notions of energy, such as the one associated to
an observer at rest outside the black hole. What would then be the new notion of energy
associated to the black hole in order to keep the energy balance? Again, the isolated horizons
formalism allows us to consider this possibility and yields a modified notion of black hole
energy. As it has been shown, this energy approaches asymptotically the horizon area for
large black holes but has some nontrivial dependence in general. We briefly comment on
this choice that has been recently advocated in the literature\new{, and show that it is related to the standard one by a Tolman redshift factor}.

The structure of this note is as follows. In Sec.~\ref{sec:stationary} we pose the problem to be considered,
namely the energy of test particles on stationary spacetimes. Sec.~\ref{sec:bh} specializes to the case
of static black holes. In the first part, we consider the Schwarzschild black hole, while in
the second part we consider the Schwarzschild--de Sitter spacetime. In Sec.~\ref{sec:bhenergy} we consider
the case where the particle falls into the black hole and how this affects the energy of the
black hole. This will depend on the notion of energy for the horizon. We consider two
cases, namely the standard choice within the isolated horizons formalism and, in Sec.~\ref{sec:rest}, a
new proposal to describe the situation from the perspective of an observer at rest outside
the horizon. We end with a discussion in Sec.~\ref{sec:discussion}.

Since this manuscript is intended to review and revisit some simple notions in general
relativity, it should be seen as a complement to the standard texts on the subject. We
have tried to make the manuscript self-contained, so the reader is not expected to possess
a very deep understanding of general relativity. Throughout the manuscript we use the
`natural' units for general relativity, namely geometrical units where $G=c=1$, and we adopt
the abstract index notation as in~\cite{wald}.

\section{ENERGY ON STATIONARY SPACETIMES}
\label{sec:stationary}

Let us start by considering an asymptotically flat spacetime $(M,g)$ with a time-like Killing
vector field (KVF) $\xi^a$, such that, in the (space-like) asymptotic region $\xi_a\xi^a=-1$. Thus, $\xi$
approaches the four-velocity of an inertial observer at infinity, and can also be interpreted as
generator of asymptotic time translations. The existence of the generator of time isometries
on the background space-time implies that there are some conserved quantities. In ordinary
`Newtonian' dynamics, time invariance implies conservation of energy, and the situation is
no different in general relativity. Here we shall focus on test particles with a four-velocity
$u^a$. The corresponding four momentum is $p^a=mu^a$, with $m$ the \emph{rest mass} of the object
(defined by $m^2:=-p_ap^a$). The quantity that is conserved, if the test particle is in free fall,
is given by,
\begin{equation}
\E=-m\,\xi_a\,u^a \label{eq:calE}
\end{equation}
(The proof is rather direct: $u^b\nabla_b\E=-m(u^bu^a\nabla_b\xi_a+\new{\xi_a\,u^b\nabla_bu^a})=0$. The first term
vanishes because $\xi$ is a KVF and satisfies the Killing equation $\nabla_{(a}\xi_{b)}=0$. The second term
is equal to zero since $u^a$ satisfies the geodesic equation $u^a\nabla_au^b=0$.) What is the physical
interpretation of the quantity $\E$? Can we say it corresponds to the energy of the particle?
The energy as measured by whom? Before answering these questions, let us recall what is the
notion of energy in special relativity. In that case one possesses a family of preferred \emph{global}
inertial observers. Their corresponding four velocities $\xi_o$ \emph{are} Killing vector fields, that are
also normalized $\xi_{oa}\,\xi_o^a=-1$. If we have a particle with four-momentum $p^a$, not necessarily
following a geodesic, then the energy of the particle, as measured by the observer $\xi_o$ is given
by,
\begin{equation}
E=-\xi_{oa}\,p^a \label{eq:ESR}
\end{equation}
This is the standard expression used in textbooks. If one introduces inertial coordinates
such that $\xi_o^a=(\partial/\partial T)^a$, then $E=p_0$, the `time component' of the four-vector $p_a$. This
quantity is also helpful to measure the speed $v$ of the particle as measured by $\xi_o$. We have
that,
\begin{equation}
E=m\,\gamma \label{eq:Emgamma}
\end{equation}
where $m^2=-p_ap^a$ and $\gamma=1/\sqrt{1-v^2}$. Then, it is straightforward to see that $E$ is the
relativistic generalization of the \emph{Kinetic energy}. If the object is at rest with respect to the
inertial observer, then $E=m$, the rest mass. If the object moves with a small velocity
$v\ll1$ (recall that we are setting $c=1$) then,
\[
E\approx m\left(1+\frac12v^2\right)=m+\frac12mv^2
\]
That is, $E$ represents the total energy of the particle as measured by the inertial observer,
that in the small velocity regime is given by the rest energy plus the Newtonian kinetic
energy. Let us denote the quantity $E$ as the \emph{generalized kinetic energy} (GKE) of the particle.
Clearly, $E$ is conserved if the particle follows a geodesic, namely if it travels at constant speed
with respect to the inertial observer $\xi_o$.

Can we generalize the energy $E$ to general relativity? The answer is simple: Yes. The
difference is that now, we do not have global inertial observers, so one has to instead consider
\emph{locally inertial observers}. Thus, at a point $p$ of the space-time $M$, a local observer $w^a$, such
that $w_aw^a=-1$, will measure the energy of a particle with four-momentum $p^a$ to be,
\begin{equation}
E=-w_a\,p^a \label{eq:Elocal}
\end{equation}
The interpretation is clear. The observer $w^a$ should be thought of as being instantly at rest,
and the quantity $E$ is the energy of the moving object as measured by that instantly inertial
observer. This quantity can always be defined for any observer and any particle, without
any restriction on the background space-time nor the dynamics followed by the observer or
the test object. The problem with this quantity is that it gives us no information about the
gravitational field. It can not, since one is constructing it by exploiting the fact that locally
any point on space-time `behaves' as if it were in special relativity. Thus, the equivalence
principle forbids us to draw any information about the gravitational field from this quantity.
What one effectively is doing is to go to an instantaneous inertial reference frame where the
gravitational field `disappears', to measure the energy of the moving object.

Let us now return to the quantity $\E$ defined for stationary spacetimes. Can it represent
the energy as measured by some observer? The answer is indeed in the affirmative. Let
us now see how that comes about. Suppose that the test object starts `at rest' at spatial
infinity, then the observer at infinity that follows the \new{world-lines} of the KVF $\xi$ will measure
the energy of the particle to be $E=-p_a\xi^a$. But notice that this is precisely the conserved
quantity $\E$. Then, this quantity corresponds to the generalized kinetic energy as measured by
an observer at rest at infinity. But, can we assign to it a further interpretation that `knows'
about the gravitational field? That is, can one define a notion of \emph{gravitational potential
energy} and relate it to $\E$? In order to answer that question, let us briefly recall the notion
of gravitational potential energy in Newtonian theory. For simplicity, consider the case of a
central object of mass $M$. The force on an object of mass $m$ at a distance $r$ is $Mm/r^2$ (recall
that $G=c=1$). Thus, the gravitational potential energy is given by $U=-Mm/r+C$
with $C$ a constant. The standard choice is to take $C=0$ such that $U(r=\infty)=0$. The
total, conserved, energy of the particle is then,
\begin{equation}
E_{\rm Tot}={\rm KE}+U=\frac12mv^2-\frac{Mm}{r}+C' \label{eq:Newton}
\end{equation}
with $C'$ a constant. Again, the standard choice is to take $C'=0$ but, in order to make contact
with general relativity, one can very well take $C'=m$. With this choice, the total Newtonian
energy of a particle at rest at infinity is $E_{\rm Tot}=m$. Obviously, the conserved quantity in
the Newtonian theory has a contribution from the gravitational field. It is then natural to
assume that in the general relativistic setting, if there is a conserved quantity representing
energy of the particle, it should contain information about the gravitational field. Thus, the
object $\E$ is the natural quantity to represent the \emph{total energy} of the particle. And, from our
previous discussion, it should have information about the (stationary) gravitational field.

Let us then define $\E$ as the total energy of the particle and study some of its properties.
The first thing to notice is that there is a simple relation with the generalized kinetic energy
which comes from the normalization of the KVF $\xi^a$. The four-velocity $w^a$ of an observer
that follows the \new{world-lines} of the KVF $\xi$ is given by,
\[
w^a=\frac{\xi^a}{V}
\]
with $V^2=|\xi_a\xi^a|$, the so-called `redshift factor'. With this, we have
\begin{equation}
\E=-p_a\xi^a=-V\,p_aw^a=V E \label{eq:EVE}
\end{equation}
Note that one can formally define a general relativistic generalization of the potential energy
by setting $\E:=E+\U$. Thus, the quantity $\U$ becomes,
\begin{equation}
\U=E\,(V-1) \label{eq:U}
\end{equation}
Or $\U=\E(1-1/V)$. With this choice, a particle that is at rest at infinity has a total energy
of $\E=m$ equal to the GKE $E$. Thus, in that case the generalized potential energy at \new{infinity}
vanishes, as in the Newtonian case. For the quantity $\U$ to capture the intuitive notion of
a \emph{negative} gravitational potential energy, the norm of the KVF should satisfy $V<1$.
Note also that, if the particle is moving when it reaches infinity, then $\E>m$. If $\E<m$
then the object can not reach infinity and is therefore bounded. A particular case of this
situation is when the particle is at rest with respect to the observer following the orbits
of $\xi^a$. In this case, the four-momentum is $p_a=mw_a$, and the corresponding energies are:
$E_\res=m$, of course, and
\begin{equation}
\E_\res=-m\xi^aw_a=-mVw^aw_a=mV\,. \label{eq:Eres}
\end{equation}
The potential energy becomes then,
\[
\U_\res=m(V-1)
\]
\new{It is worth noting a feature of $\U$ that has no Newtonian counterpart. For a given position, that is, for a given value of $V$, the potential energy $\U=E(V-1)$ is proportional to the GKE $E=m\gamma$, so that it depends on the state of motion of the particle and not only on its position. This reflects the fact that in general relativity gravity couples to the total energy of an object, kinetic energy included, and not only to its rest mass. For a particle at rest, $\U_\res$ depends only on position; this is the case for which the properties discussed in the next section are established.}

Let us now consider the particular case of interest, namely when the spacetime corresponds
to the exterior of an eternal black hole.

\section{ENERGY IN THE VICINITY OF A BLACK HOLE}
\label{sec:bh}

In this section we shall concentrate our attention on the different definitions of energy
for test particles in the vicinity of stationary black holes. This section has two parts. In the
first one, we discuss the asymptotically flat case, where for simplicity we focus our attention
on the Schwarzschild case. In the second part, we consider a non-vanishing and positive
cosmological constant, where the notion of an asymptotic inertial observer does not exist.

\subsection{\texorpdfstring{Asymptotically flat case ($\Lambda=0$)}{Asymptotically flat case (Lambda=0)}}
\label{sec:flat}

For simplicity and without loss of generality, the spacetime we are going to consider is
given by the Schwarzschild metric,
\begin{equation}
\dd s^2=-\left(1-\frac{2M}{r}\right)\dd t^2+\left(1-\frac{2M}{r}\right)^{-1}\dd r^2+r^2\dd\Omega^2 \label{eq:schw}
\end{equation}
From this expression we see that the timelike KVF is $\xi^a=(\partial/\partial t)^a$, with norm given by,
\[
V=|\xi_a\xi^a|^{1/2}=\left(1-\frac{2M}{r}\right)^{1/2}
\]
Recall that this spacetime is asymptotically flat, and has a horizon at $r_{\rm h}=2M$, where $M$
is the Arnowitt--Deser--Misner (ADM) mass of the spacetime and one can also associate it to
the mass of the black hole (more about this later). Note that in the exterior of the black
hole, namely when $r>2M$, the red-shift factor satisfies $V<1$ and $V\to0$ as one approaches
the horizon. Let us now analyse the other quantities we have defined. First, let us write the
relation between the GKE $E$, as measured by an observer at rest with respect to the BH at
a `distance' $r$, and the total, conserved energy $\E$,
\begin{equation}
E=\frac{\E}{\left(1-\frac{2M}{r}\right)^{1/2}} \label{eq:EofcalE}
\end{equation}
From here we can use the relation $E=m\gamma$ to find the speed of the particle with total energy
$\E$, as measured by the observer at rest at a `distance' $r$,
\begin{equation}
v=\left[1-\left(\frac{m}{\E}\right)^2\left(1-\frac{2M}{r}\right)\right]^{1/2}. \label{eq:speed}
\end{equation}
From (\ref{eq:EofcalE}) one can see that, for a particle with given energy $\E$, the GKE as measured by
observers at rest will increase and diverge as one approaches the horizon. Eq.~(\ref{eq:speed}) is the
generalization of the Newtonian equation relating speed with the total energy and distance
from the origin.\footnote{Recall that, in the Newtonian case, for a particle of total energy $E_n$ as defined in the previous section
in Eq.~(\ref{eq:Newton}), the speed $v_n$ is given by $v_n=\sqrt{2\left(\frac{E_n}{m}-1\right)+\frac{2M}{r}}$. Note that this and the GR expression are
rather different, except for the case when $\E=m$ in general relativity and $E_n=m$ in the Newtonian
theory, where both theories yield $v=\sqrt{2M/r}$. Incidentally, this `coincidence' is precisely what allowed
Michell, with his 1783 calculation within Newtonian mechanics, to predict the ``correct'' value for the
black hole radius. We see that in both GR and Newtonian dynamics, for an object of mass $M$ and radius
$r$, the escape velocity is equal to 1, precisely when $r=2M$.}
Let us see what information we can extract from it. First, suppose the
particle has an energy $\E>m$, then it can reach infinity and there, the speed is given by
$v|_\infty=\sqrt{1-(m/\E)^2}$. If the total energy is less than the rest mass, $\E<m$, then the particle
can not reach infinity. For instance if it is `thrown' upward and then follows a radial geodesic,
then the maximal radial distance it will attain is,
\begin{equation}
r_{\max}=\frac{2M}{1-\left(\frac{\E}{m}\right)^2} \label{eq:rmax}
\end{equation}
before falling again in the direction of the black hole. We can compare this quantity with the
corresponding Newtonian quantity $r^*_{\max}$ given by: $r^*_{\max}=\frac{M}{1-\frac{E_n}{m}}$, that relates the maximum
distance that a test particle would travel away from a point (or spherical) object of mass $M$
(located at $r=0$), as a function of the energy $E_n$ of the test particle. Finally, note that if
we throw the particle in the direction of the black hole, its speed will approach $v\to1$ as it
reaches the horizon, independently of the value of $\E$.

\medskip
\noindent\emph{Accelerated motion `at rest'.} Let us consider the following situation. We start with a test
particle of rest mass $m$ following the orbit of the KVF $\xi^a$ at infinity. The total energy is
clearly $\E=m$. Now, let us assume that we lower the particle in a quasi-static fashion from
infinity to a distance $r$, in such a way that it is at rest.\footnote{We do not need to perform such a procedure, namely to lower the particle in a quasi-static fashion. We
could as well consider, as initial condition, a test particle at rest at a distance $r$.}
We know that in the Newtonian
theory, the energy of the particle decreases, since it only possesses potential energy. We
expect that in general relativity the total energy $\E$ also decreases. In fact, from Eq.~(\ref{eq:Eres}) we
see that the total energy is given by
\begin{equation}
\E_\res=m\sqrt{1-\frac{2M}{r}} \label{eq:Eres-schw}
\end{equation}
There are two points to remark. As expected, the energy decreases as $r$ decreases but as
the particle approaches the horizon, the total energy approaches zero. That is, a particle
that is held by an observer at rest just outside the black hole horizon has an energy $E=m$,
as measured by the observer, but its total energy is much smaller, approaching zero at the
horizon. That is, it has lost all its energy due to the negative gravitational potential energy
given by,
\begin{equation}
\U_\res=m\left[\left(1-\frac{2M}{r}\right)^{1/2}-1\right], \label{eq:Ures-schw}
\end{equation}
which takes values in the interval $[-m,0]$. Note also that there is a ``positive energy result''
at hand: The conserved energy of a test particle `at rest' is strictly positive in the region
where it can be defined (recall that in the interior region, the notion of energy for the
conserved quantity is no longer valid, since the Killing vector field $\xi$ becomes spacelike).
\new{This is precisely the energetics behind the Geroch process~\cite{geroch,bekenstein}: if the particle is lowered quasi-statically from infinity by means of an idealized rope, the energy $m-\E_\res=-\U_\res=m(1-V)$ is delivered to the agent at infinity, and it approaches the full rest energy $m$ as the particle approaches the horizon. We shall return to this in Sec.~\ref{sec:bhenergy}, when the particle is dropped into the black hole.}

The second observation is that this potential energy (\ref{eq:Ures-schw}) satisfies several desired properties. In particular:\\
i) For $r\gg2M$, the quantity approaches $\U_\res\approx-Mm/r$.\\
ii) It is always negative, and\\
iii) It serves as a potential for the force felt by the test particle to be held at rest. The
magnitude of the force is $F=mg$ with $g$ the norm of the acceleration $\alpha^b=u^a\nabla_au^b$. That
is, $g=\sqrt{|\alpha_b\alpha^b|}$. In our case,
\begin{equation}
F=\frac{mM}{r^2}\,\frac{1}{\left(1-\frac{2M}{r}\right)^{1/2}}=\partial_r\,\U_\res \label{eq:force}
\end{equation}
\new{Here the derivative is taken with respect to the areal radius $r$. The invariant form of this statement uses instead the proper radial distance $s$, with $\dd s=V^{-1}\dd r$:
\begin{equation}
\frac{\dd\,\U_\res}{\dd s}=V F=\frac{mM}{r^2}\,. \label{eq:force-proper}
\end{equation}
The right-hand side is the redshifted force, namely the force that an observer at infinity must exert, through an idealized massless rope, to hold the particle at rest. Its work along a proper displacement equals the change in the conserved energy, $\dd\E_\res=VF\,\dd s$, which is why (\ref{eq:force-proper}), rather than (\ref{eq:force}), is the relation directly tied to energy balance. Note also that $VF/m=Vg\to\kappa=1/4M$ as $r\to2M$: the surface gravity, which will play a central role in Secs.~\ref{sec:bhenergy} and~\ref{sec:rest}, is precisely the horizon limit of this redshifted force per unit mass.}

Note that the potential energy $\U_\res$ that we have here defined improves on the proposal put
forward in~\cite{woodhouse}, which does not satisfy the properties i)--iii) above. \new{While the Killing energy $mV$ of a static particle that underlies (\ref{eq:Ures-schw}) is classical, as the discussion of the Geroch process above makes clear, to the best of our knowledge its interpretation as a gravitational potential energy satisfying i)--iii) has not been emphasized elsewhere.}

\medskip
\noindent\emph{Circular motion.} Let us now consider the test particle of rest mass $m$ following a stable
circular orbit. When there is angular momentum, the radial coordinate satisfies a potential
equation of the form,
\[
(\dd r/\dd\tau)^2+\new{V_{\rm eff}}(r)=(\E/m)^2-1\,,
\]
with $\new{V_{\rm eff}}(r)=\new{-2M/r}+L^2/(m^2r^2)-2ML^2/(\new{m^2}r^3)$ and $L,\E$ the conserved quantities. Circular
orbits are determined by the extrema of the potential. It has two extrema, one maximum
and one local minimum, that represents stable circular orbits. The smallest such orbit is
when both extrema coincide, which occurs when $L^2=12M^2m^2$. Then, the value of this
minimum radius $r_c^-$ is equal to $r_c^-=6M=3r_{\rm h}$. If one evaluates the potential $\new{V_{\rm eff}}(r)$ for this
value, one finds that,
\begin{equation}
\E_c^-=m\,\frac{2\sqrt2}{3} \label{eq:isco}
\end{equation}
Let us now consider a test particle that is on a circular orbit `at infinity' for which $\E_\infty=m$
(as could be expected\footnote{Note that, just as in the Newtonian theory, the speed $v$ of circular orbits decreases as the radius of the orbit
increases so that, in the $r\to\infty$ limit, the speed vanishes, and therefore, the total energy $\E$ corresponds
to the rest mass of the object.}).
Let us now assume that the particle \new{loses} energy as it orbits the
black hole in such a way that it follows quasi-circular orbits. It is immediate to see that it
will, at most, \new{lose} an energy equal to $\Delta\E=\E_\infty-\E_c=m(1-2\sqrt2/3)\approx0.057\,m$, before
plunging into the black hole. This is clearly the energy source of accretion disks around
black holes, that lose energy by radiating in the electromagnetic spectrum. One can also
find the speed of the particle on the last stable circular orbit to be,
\[
v_c^-=\left[1-\left(\frac{3}{2\sqrt2}\right)^2\left(1-\frac13\right)\right]^{1/2}=\frac12\,.
\]
That is, the top speed of a test particle on a \new{stable} circular orbit is precisely 50\% of the speed of
light. \new{On the unstable circular orbits with $3M<r<6M$, the speed measured by static observers, $v=[M/(r-2M)]^{1/2}$, keeps growing and reaches the speed of light at the photon sphere $r=3M$.}

We end here our discussion of the energetics of test particles on an asymptotically flat
static spacetime. Clearly, most of the results described here will be qualitatively similar for
other asymptotically flat static black holes. As should be clear from our previous discussion,
and as we have argued in this part, the right notion of energy to consider is $\E$, if one
wants to capture the contribution to the energy coming from the gravitational field. Let us
now explore the situation when the cosmological constant is non-vanishing and we have no
asymptotically flat regions.

\subsection{\texorpdfstring{de Sitter Black Holes ($\Lambda>0$)}{de Sitter Black Holes (Lambda>0)}}
\label{sec:dS}

So far we have focused on test particles that live on static patches of asymptotically flat
spacetimes and have focused on the notion of energy for them. The existence of asymptotic
observers has proven crucial for several of our discussions. In particular, it allows us to normalize the KVF $\xi$ such that its norm becomes unity at infinity. The physical interpretation
is that this preferred observer is the one for which the conserved quantity $\E$ coincides with
the GKE $E$, so that we can interpret $\E$ as the (GKE) energy as measured by an inertial
observer at infinity. When we depart from the $\Lambda=0$ case, the first observation is that we
no longer have an asymptotically flat region. In the patches where the metric looks static,
the vector $\partial/\partial t$ is not canonically normalized, so we immediately \new{lose} the interpretation we
had in the flat case. Thus, we need to adopt a slightly different strategy. In the remainder
of this part, we shall consider the de Sitter case with a positive cosmological constant.

Let us start by considering the simplest static metric in this case, the so-called
Schwarzschild--de Sitter (SdS) metric:
\begin{equation}
\dd s^2=-f^2\dd t^2+f^{-2}\dd r^2+r^2\dd\Omega^2 \label{eq:sds}
\end{equation}
where $f^2=\left(1-\frac{2\mu}{r}-\frac{r^2}{\ell^2}\right)$, with $\ell=\sqrt{3/\Lambda}$ the natural length scale defined by the cosmological
constant, and $\mu$ the `mass parameter'.\footnote{While $\mu$ is sometimes regarded as the mass of the black hole, this assumption is unsupported as was
discussed in detail in~\cite{cg}.}
The function $f^2$ has two zeros, corresponding to
the two horizons. The smallest root $r_b$ corresponds to the black hole horizon, while the
larger one $r_c$ defines the cosmological event horizon. The static patch corresponds to the
region $r_b<r<r_c$. One thing to note is that in no place does $f^2=1$, which means that
$\bar\xi^a=(\partial/\partial t)^a$ can never correspond to the four velocity of a static observer. One needs to
normalize it appropriately. Second, note that from all the possible static observers, there is
one (sphere worth of) \new{observer(s)} that is special. Let us see now \new{how} to select those
observers. If a static observer is close to the black hole horizon then, in order to remain
static, she needs to accelerate away from the black hole. On the other hand, if the observer is
now close to the cosmological horizon, in order to remain static she would need to accelerate
\emph{towards} the black hole due to the cosmic expansion. Then, there is an intermediate point
$r_i$ where a static observer is inertial; it is an unstable `equilibrium point'. If a particle
is following the KVF orbit there and one pushes it slightly towards the BH it will fall to it,
while if the push is in the outward direction, then the particle `falls' toward the cosmological
horizon. The natural normalization for the \new{KVF} presents itself before our eyes: \new{choose} the
KVF $\xi$ such that, for the inertial observer at $r=r_i$, we have the normalization $(\xi\cdot\xi)|_{r_i}=-1$.
The value of this preferred radius is given by $r_i=(\mu\ell^2)^{1/3}$. Thus, we have,
\begin{equation}
\xi^a=\frac{\bar\xi^a}{f(r_i)}=\frac{1}{\sqrt{1-3\left(\frac{\mu}{\ell}\right)^{2/3}}}\left(\frac{\partial}{\partial t}\right)^a. \label{eq:xi-sds}
\end{equation}
\new{(The normalization (\ref{eq:xi-sds}) requires $27\mu^2<\ell^2$, i.e., $9\Lambda\mu^2<1$, which is precisely the condition for the black hole and cosmological horizons to be distinct. In the Nariai limit $27\mu^2=\ell^2$ the radii $r_b$, $r_i$ and $r_c$ all coalesce at $r=3\mu$.)}
With this choice, the red-shift factor $V$ takes the form:
\[
V=\frac{f(r)}{f(r_i)}=\frac{1}{\sqrt{1-3\left(\frac{\mu}{\ell}\right)^{2/3}}}\left(1-\frac{2\mu}{r}-\frac{r^2}{\ell^2}\right)^{1/2}.
\]
One could argue that these inertial observers play the same role as the asymptotic observers
do for the $\Lambda=0$ case, since they are both inertial and `far away' (if $\ell\gg\mu$) from the black
hole. Thus, one can now define the conserved quantity $\E$ as before: $\E=-p_a\xi^a=VE$.

We can again consider the situation in which we have radial geodesics, in which case the
relation between speed and `radius' $r$, for a test particle of total energy $\E$ is given by,
\begin{equation}
v=\sqrt{1-\left(\frac{m}{\E}\right)^2V^2} \label{eq:speed-sds}
\end{equation}
We can follow much of the analysis that we did in the previous part for the asymptotically
flat case. Note that the role that infinity played in the previous case is now played by the
`point $r_i$'. Let us now revisit some of the features that resemble the flat case. The first
observation is that a test particle with $\E=m$ is necessarily at rest at $r_i$, since $V(r_i)=1$. If
$\E<m$ then the particle is either in a region $r<r_{\max}<r_i$ closer to the black hole, or `on
the other side of the potential', namely $r>r_{\min}>r_i$. Here $r_{\max}$ represents the maximum
value that $r$ can attain in the region $[r_b,r_i]$, and $r_{\min}$ is the minimum value $r$ can take in the
region $[r_i,r_c]$. They are given by the real positive roots of the equation
\[
\left(1-\frac{2\mu}{r}-\frac{r^2}{\ell^2}\right)=\new{\left(\frac{\E}{m}\right)^2}\left(1-3\left(\frac{\mu}{\ell}\right)^{2/3}\right).
\]
Clearly, if $\E>m$ the particle can climb up the potential away from the black hole and \new{cross the cosmological horizon}. Conversely, if it is originally in the cosmological region ($r>r_i$), and thrown to the
center of the spacetime it will overcome the cosmic repulsion and reach the region where the
attraction of the black hole dominates, and fall into it.

It is perhaps illustrative to estimate the value of the radius $r_i$ where the transition occurs,
for the value of the cosmological constant as recently measured. Its value is currently
estimated to be \new{$\Lambda\approx1.1\times10^{-52}\,{\rm m}^{-2}$~\cite{planck}. Since $r_i=(\mu\ell^2)^{1/3}=(3\mu/\Lambda)^{1/3}$, the transition radius scales as $\mu^{1/3}$.} Let us now estimate $r_i$ for three different values of
$\mu$. First, let us consider a stellar mass black hole with mass of the order of the mass of the
sun. In that case \new{$\mu\approx1.5\times10^3\,{\rm m}$, and $r_i\approx3.4\times10^{18}\,{\rm m}\approx110$ pc}. Note that the Schwarzschild--de Sitter metric describes not only a black hole, but the exterior region of a spherically
symmetric non-rotating star. Thus, we see that for a system such as the solar system, the
cosmic repulsion dominates at scales that are well within the galaxy.\footnote{Of course, we do not see this effect inside our galaxy, since the contribution from other stars and the
galactic centre outweighs that from the cosmic repulsion.}
Next, let us consider
the case of a super-massive black hole at the galactic center with mass of the order of $10^6$
solar masses. In that case, \new{$r_i\approx11$ kpc}, which is again smaller than the galactic radius.
Finally, for a mass of the order of the Milky Way galaxy\new{, $\sim10^{12}$ solar masses}, we have \new{$r_i\approx1.1$ Mpc}. Recall that
the closest \new{large} galaxy to ours is Andromeda at about \new{0.77} Mpc, so it is very close to the transition
point.\footnote{But in that case, since Andromeda has a mass comparable to that of our galaxy, the test mass approximation clearly fails and one would have to treat the system as a two body problem, in which case the
effect of the cosmological constant is less transparent.}

Let us return to the radial geodesic we were considering. Suppose we have a test particle
and we throw it radially towards $r_i$, with just enough speed so that $\E=m$. From the
previous considerations we know that the particle will have zero speed upon arrival to the
point $r=r_i$. Let us now calculate the proper time that it would take to reach that point.
As usual one starts from the `first integral' of the geodesic equations: $g_{ab}u^au^b=-1$, and
by using the spherical coordinates and considering moving in only the radial direction, one
arrives at the expression, for generic $\E$,
\begin{equation}
\dd\tau=\frac{\dd r}{\sqrt{\left(1-3\left(\frac{\mu}{\ell}\right)^{2/3}\right)\left(\frac{\E}{m}\right)^2-1+\frac{2\mu}{r}+\frac{r^2}{\ell^2}}} \label{eq:dtau}
\end{equation}
Now, we take the case $\E=m$ that we are considering and by redefining the radial coordinate
to be $\tilde r:=r/r_i=r/(\mu\ell^2)^{1/3}$, the proper time $\tau$ to go from the point $r_0$ to $r_i$ is given by the
expression
\begin{equation}
\tau=\ell\int_{\tilde r_0}^1\frac{\dd\tilde r}{\sqrt{-3+\frac{2}{\tilde r}+\tilde r^2}}\,, \label{eq:tau}
\end{equation}
that diverges. Also, note that if the conserved energy is slightly different from $m$, say
$0<\E-m\ll m$, then the corresponding integral converges, so one can transition from the
black hole region to the cosmological region in a finite proper time.\footnote{And if the energy is smaller than $m$, one reaches the turning point at a finite proper time.}

\medskip
\noindent\emph{Accelerated motion at rest.} Let us consider the following situation. We start with a test
particle of rest mass $m$ following the orbit of the KVF $\xi^a$ at the point $r_i$. The total energy
is clearly $\E=m$. Now, let us assume that we lower the particle in a quasi-static fashion
in such a way that it is at rest at a distance $r$. We know from the previous part that the
energy $\E$ of the particle decreases, since it only possesses potential energy. From Eq.~(\ref{eq:Eres}) we
see that the total energy is given by
\begin{equation}
\E_\res=\frac{m}{\sqrt{1-\new{3\left(\frac{\mu}{\ell}\right)^{2/3}}}}\left(1-\frac{2\mu}{r}-\frac{r^2}{\ell^2}\right)^{1/2} \label{eq:Eres-sds}
\end{equation}
There are two points to remark. As expected, the energy decreases as $r$ decreases but as
the particle approaches the horizon, the total energy approaches zero. That is, a particle
that is held by an observer at rest just outside the black hole horizon has an energy $E=m$,
as measured by the observer, but its total energy is much smaller, approaching zero at the
horizon. That is, it has lost all its energy due to the negative gravitational energy given by,
\begin{equation}
\U_\res=m\left[\left(\frac{1-\frac{2\mu}{r}-\frac{r^2}{\ell^2}}{1-3\left(\frac{\mu}{\ell}\right)^{2/3}}\right)^{1/2}-1\right], \label{eq:Ures-sds}
\end{equation}
which, as in the previous case, takes values in the interval $[-m,0]$. Note that in this case
we also have a positive energy result, given that the conserved energy of a test particle is
strictly positive in the region where it can be defined (recall that in the interior regions, the
notion of energy for the conserved quantity is no longer valid, since the Killing vector ceases
to be timelike). The potential energy takes its minimum value $-m$ at both horizons and is
zero at the point $r_i$.

\medskip
\noindent\emph{$\Lambda<0$ Case.} Let us now briefly comment on the other possibility when there is a
cosmological constant present, namely the case of asymptotically AdS spacetimes for $\Lambda<0$.
This case has drawn a lot of attention due to the conjectured AdS/CFT dualities. The first
observation is that in static coordinates for which the metric takes the form,
\begin{equation}
\dd s^2=-\left(1-\frac{2\mu}{r}+\frac{r^2}{\ell^2}\right)\dd t^2+\left(1-\frac{2\mu}{r}+\frac{r^2}{\ell^2}\right)^{-1}\dd r^2+r^2\dd\Omega^2 \label{eq:sads}
\end{equation}
we do not have, as in the de Sitter case, a preferred asymptotic observer, since the
norm $V$ of the Killing field $\xi$ tends to infinity as $r$ grows. There is however a preferred
observer located at the point $r_m=(2\mu\ell^2)^{1/3}$, where the norm of the KVF is unit, $V=1$.
Thus, the total energy $\E$ as defined by $\xi$ has the interpretation of being the energy as
measured by a static observer at $r=r_m$. The main difference with the previous cases is
that this observer is not invariantly defined (it depends on the choice of coordinate $t$), in
contrast to the previous two cases where the preferred static observers follow geodesics. As
is well known for the AdS case, there is a very large `potential wall' for large values of $r$,
such that one has to \new{impart an} increasing amount of energy to a radial particle to climb
the wall. In order to send it to infinity, the energy would also have to be unbounded. Thus,
no physical (massive) object can reach infinity $\mathcal{I}$. \new{Note also that, since there is no static observer with unit-norm four-velocity near infinity, $\mu$ cannot be interpreted as the energy of the black hole as measured by such an observer. This does not contradict the standard results on conserved charges: the charge associated with the Killing field $\partial/\partial t$ at the conformal boundary, as defined by Ashtekar and Magnon (see also Ashtekar and Das)~\cite{amd} or by Henneaux and Teitelboim~\cite{ht}, is precisely $\mu$. The subtlety lies only in the operational interpretation of this charge in terms of observers.}

So far, we have considered test particles in a static black hole background and studied
the different types of energy one can consider for the particle. What we have not considered
is what happens when the particle crosses the horizon and falls into the black hole. That
will be the subject of the next section.

\section{ENERGY OF THE BLACK HOLE}
\label{sec:bhenergy}

In this section we shall consider the issue of the energy of the particle and how it relates
to the energy of the black hole. In the previous section we considered only test particles,
where the test object does not affect the background black hole. In this part we shall relax
that assumption a bit. We will still neglect any back-reaction on the background geometry
as long as the particle remains outside the horizon, but shall consider the change in area
and mass of the black hole as the objects fall into the horizon.

Let us start with the simplest case, namely an object of rest mass $m$ at rest at infinity
in an asymptotically flat spacetime of a black hole of mass $M$. Here, the total energy of
the spacetime, as captured by the ADM mass, is then $M_{\rm ADM}=M+m$. We are assuming
that $m\ll M$ and therefore, there is little `interaction' between the black hole and the test
object. Here, the mass $M$ of the black hole refers to the horizon mass $M_{\rm h}$ as defined by the
\emph{isolated horizon} formalism~\cite{ih}.\footnote{It has been shown that for asymptotically flat stationary electro-vac spacetimes the horizon mass and
the ADM mass coincide. Here we are assuming that the presence of the test object does not modify the
horizon mass $M$, so our approximation is valid.}

Let us now assume that we drop the object radially towards the BH, following a radial
geodesic. Then the total energy of the particle is $\E=m$, and that quantity shall be
conserved along the geodesic. The object will cross the horizon at a finite proper time
(assuming we did not actually drop it from infinity but rather from a sufficiently distant
asymptotic region). The horizon will react to the crossing of the object and will quickly
settle down to a new isolated horizon. The question now is: what is the new horizon mass
$M'_{\rm h}$? To answer that question is rather simple. After the object falls into the black hole
and it settles, to the future the spacetime will be vacuum and approximately static (in our
approximation we are assuming the energy radiated is negligible with respect to $m$), so it
is very well approximated by a Schwarzschild metric with mass $M'_{\rm h}$. In the isolated horizon
formalism it has been established that for static spacetimes with an isolated horizon, as
is our case, the ADM mass and the horizon mass coincide (more generally, when there is
radiation, the Bondi energy at future null infinity and the `mass at $i^+$' coincide~\cite{ih}). That
is, the new horizon mass will simply be,
\begin{equation}
M'_{\rm h}=M_{\rm h}+\E \label{eq:Mnew}
\end{equation}
which means that the change in horizon mass is equal to the total energy of the test object,
$\delta M_{\rm h}=\E$. The next scenario is to consider a particle that has a total energy smaller \new{than}
its rest mass, $\E<m$. This could be the case if initially the test object is at rest at some
distance from the black hole and is then released to fall into the black hole following a radial
geodesic. Just as in the previous case, the horizon mass $M_{\rm h}$ of the black hole will change by
an amount equal to the total energy of the test object $\E$. For our case here,
\begin{equation}
\delta M_{\rm h}=m\sqrt{1-\frac{2M}{r_\res}}\,, \label{eq:dMh}
\end{equation}
where $r_\res$ is the distance from which the object is released. The extreme case is when the
object of mass $m$ is held at rest just outside the horizon. As we saw in the previous section,
the total energy of the object can be arbitrarily small as one approaches the horizon, so when
released into the black hole, the change in BH mass will be arbitrarily small, approaching
zero in the limit.

\new{This is the Geroch process~\cite{geroch}: together with the energy $m(1-V)$ extracted at infinity while lowering the particle (Sec.~\ref{sec:flat}), the total energy $m$ is accounted for, and in the limit the whole rest mass is converted into work at infinity while the black hole mass, and hence its area, remains unchanged. The limit is, however, an idealization valid for point particles. For an object of proper radius $b$, the center of mass cannot be held at a proper distance smaller than $b$ from the horizon. Since near the horizon $V\approx\kappa s$, with $s$ the proper distance to the horizon and $\kappa$ the surface gravity introduced below, this gives $\delta M_{\rm h}\gtrsim\kappa\,m\,b$ and, by the first law (\ref{eq:firstlaw}), $\delta A_{\rm h}\gtrsim8\pi\,m\,b$, as first pointed out by Bekenstein~\cite{bekenstein}. When the quantum thermal atmosphere of the black hole is taken into account, the buoyancy it exerts on the object modifies these energetics further~\cite{uw}.}

Let us now recall the first law of black hole mechanics. For the non-rotating case we are
considering, there is a relation between the change of mass of the black hole and how much
it `grows' as measured by a change in area. The relation is given by,
\begin{equation}
\delta M_{\rm h}=\frac{\kappa}{8\pi}\,\delta A_{\rm h} \label{eq:firstlaw}
\end{equation}
where $\kappa$ is the so-called surface gravity and $A_{\rm h}$ is the horizon area $A_{\rm h}=4\pi r_{\rm h}^2$. The surface
gravity can be found from the equation $\ell^a\nabla_a\ell^b=\kappa\ell^b$ with $\ell$ the null generator of the isolated
horizon. For static spacetimes it is standard to set $\ell^a=\xi^a$ but, in principle, other choices
are allowed. What Eq.~(\ref{eq:firstlaw}) is telling us is that the horizon mass can only be a function
of the area $A$ (or radius $r$, of course). Thus $\kappa$ could be in principle an arbitrary function of
$r_{\rm h}$. For any asymptotically flat spacetime in vacuum, with a non-rotating isolated horizon,
which includes also the Schwarzschild spacetime, we have that $\kappa_o(r_{\rm h})=1/(2r_{\rm h})$, from which
one can integrate the first law to obtain~\cite{ih},
\begin{equation}
M_{\rm h}(r_{\rm h})=\frac12\int_0^{r_{\rm h}}\beta(r)\,\dd r \label{eq:Mh}
\end{equation}
with $\beta(r)=2r\,\kappa(r)$. This formula is valid for any non-rotating static horizon. Note that,
for the standard choice for $\kappa_o$, $\beta_o=1$, from which we obtain a horizon mass $M_{\rm h}(r_{\rm h})=r_{\rm h}/2$.
Note also that when integrating the first law, one is setting $M_{\rm h}(r_{\rm h})|_{r_{\rm h}=0}=0$.\footnote{This choice is modified when there are nontrivial static solitonic solutions as in EYM.}

From our previous discussion regarding energetics, we see that the standard choice for
surface gravity leading to an equality between horizon and ADM masses (for electro-vac
spacetimes) is fully justified. The reason is that, as we have seen, the change in mass of
the horizon has to be equal to the total energy $\E$ of the in-falling object. This energy is
`measured' by the KVF $\xi$, and then its natural extension to the horizon is the now null
vector $\ell=\xi$. Energy is conserved in a ``global sense''.

Let us now discuss the case of a test particle falling into a de Sitter black hole. Some of
these observations have been previously discussed in~\cite{cg}, so we refer the interested reader to
that paper for a deeper discussion of black hole mass and energy in the de Sitter context. The
important point to remember here is that the role of the asymptotic observer in AF gravity
gets replaced by the static observer located at $r_i$. The Killing vector field \new{$\bar\xi=\partial/\partial t$} was
rescaled to be of unit norm at the sphere corresponding to $r_i$, in order to define the total
energy. The same strategy was followed in~\cite{cg} to define the BH horizon mass and energy.\footnote{Which, incidentally, do \emph{not} coincide for the cosmological horizon, in the $\Lambda\neq0$ case~\cite{cg}.}
With that choice for the Killing vector field $\xi$, we recover the same global conservation of
energy we had in the asymptotically flat case, as the test particle falls into the black hole
horizon. The change in horizon mass $\delta M_{\rm h}$ corresponds exactly to the total energy $\E$ of the
test particle, as measured by $\xi$. The Bondi energy is now replaced by the energy associated
to the cosmological horizon~\cite{cg}. Energy is conserved in a global sense.

In this part we have seen that the notion of total energy $\E$ for a test particle is consistent
with the standard definition of horizon energy for the black hole in order to have (global)
conservation of energy. Note that any other choice will not benefit from this feature. Let us
now comment on one such possibility.

\section{ANOTHER PROPOSAL: ENERGY FOR OBSERVERS AT REST}
\label{sec:rest}

Any other choice of vector $\ell$ of the type $\alpha\xi$, when extended to the horizon will give
different values of the surface gravity $\kappa_\ell$ and therefore, of the measure of energy falling into
the black hole, and of its horizon mass. One possibility is that one takes, for instance, the
vector $w^a=\xi^a/V$ that, as we have seen, corresponds to the four-velocity of an observer at
rest. In that case, we can not really extend it to the horizon since $V=0$ there. What one
could do is to stay a little bit outside the horizon and define the various quantities there.
This is precisely the proposal we shall comment on now.

As we have already noted, if we choose the vector $w^a=\xi^a/V$ and try to extend it to the
horizon as a generator of the horizon, we do not succeed since the redshift factor $V$ vanishes
there. A recent proposal is to consider instead an observer that is at a small distance, of
the order of the Planck length, outside the horizon~\cite{gp}. Next, the generalized kinetic energy
is taken as the ``right'' notion of energy and then the authors consider a modified first law
that reflects this observer at rest. Let us describe this procedure and offer some comments.
The first step is to look at the first law and see if one can find a new $\bar\kappa$ that captures the
fact that the description of the particle falling will be done by the observer at rest. The first
attempt, namely to generalize the equation $\ell^a\nabla_a\ell^b=\kappa\ell^b$ (valid at the horizon) and use the
acceleration $g$ of the observer at rest given by $g=\frac{M}{r^2}\left(1-\frac{2M}{r}\right)^{-1/2}$. Note that, at the horizon,
the surface gravity $\kappa$ satisfies $\kappa=Vg$, namely it is equal to the red-shifted value of the
acceleration (that diverges on the horizon). Thus, one could propose that the ``effective''
surface gravity just outside the horizon is given by,
\begin{equation}
\bar\kappa=\frac{\kappa}{V}=\kappa\left(1-\frac{2M}{r}\right)^{-1/2} \label{eq:kappabar}
\end{equation}
Note that this is precisely the proposal of~\cite{gp}. Their argument is the following: As seen by
the observer at rest, the energy that the observer assigns to the particle is $E=\E/V$. Thus,
according to the observer, the black hole mass $\tilde M$ must have changed by $E$ and not $\E$. Then
\begin{equation}
\delta\tilde M=\frac{\E}{V}=\frac{\delta M_{\rm h}}{V} \label{eq:dMtilde}
\end{equation}
Then, using the first law, they conclude that $\delta\tilde M=\frac{\kappa}{8\pi V}\delta A$. Thus, the corresponding surface
gravity for the observer's mass $\tilde M$ is $\bar\kappa=\kappa/V$.

\new{It is worth stressing what this modification does and does not change. Since $\delta\tilde M=\delta M_{\rm h}/V$ and $\bar\kappa=\kappa/V$, the two first laws describe exactly the same change of area,
\begin{equation}
\frac{\delta\tilde M}{\bar\kappa}=\frac{\delta M_{\rm h}}{\kappa}=\frac{\delta A_{\rm h}}{8\pi}\,. \label{eq:tolman}
\end{equation}
Equivalently, $\bar\kappa/2\pi=\kappa/(2\pi V)$ is the Hawking temperature blueshifted by the Tolman factor $1/V$~\cite{tolman}, i.e., the local temperature measured by the observer at rest, and $\tilde M$ is the correspondingly blueshifted energy. Both descriptions thus assign the same entropy, $S=A_{\rm h}/4$, to the horizon. They differ only by the Tolman factor, and the question at stake is not the thermodynamics of the horizon but which quantity deserves to be called the energy of the black hole.}

With this choice, there is a local sense of conservation, which means that the energy that
the observer sees falling into the black hole is recovered in the mass of the black hole when
it grows. However, if we decide to consider this as the notion of energy of the black hole
horizon, we \new{lose} any contact with the \emph{canonical} definition of total energy at infinity, so the
energy balance that we had before with the energy $\E$ and the ADM energy for the whole
spacetime is lost. This, in particular, implies that we have to forgo the notion of `global'
conservation of energy, even in the restricted context of stationary backgrounds. It is our
belief that these considerations render the proposal of~\cite{gp} difficult to justify, from a physical
viewpoint.\footnote{Note also that in the proposal of~\cite{gp} and in subsequent papers, it is argued that the `mass' $\tilde M$ of the
black hole is proportional to its area. This is only partially true after certain approximations have been
made that, in particular, involve the large area limit. \new{Concretely, near the horizon $V\approx\kappa s$, with $s$ the proper distance from the observer to the horizon, so that $\bar\kappa\approx1/s$ and, for fixed $s$, $\tilde M\approx A/(8\pi s)$; this relation receives corrections away from the near-horizon regime.} One should, nevertheless, not forget that the actual
relation between $\tilde M$ and area is much more complicated. Making them equal and, furthermore, equating
them as operators is then an unjustified extrapolation.}

\section{DISCUSSION}
\label{sec:discussion}

In this note we have considered test particles on stationary spacetimes and, as concrete
examples, in the context of asymptotically flat and de Sitter static black hole spacetimes.
As we have argued in detail, there is an important difference between the two main notions
of energy. The direct generalization to general relativity of the energy as measured by an
observer, even when well defined independently of the spacetime geometry and observer,
does not provide us with information about the gravitational contribution to the energy. On
the other hand, when the spacetime is stationary, there is a conserved quantity that can be
interpreted as the total energy of the particle, \emph{including} the gravitational (potential energy)
contribution. We have also isolated the gravitational potential energy for certain cases and
shown that it satisfies a set of physically motivated properties\new{, and pointed out its relation to the classic Geroch process}.

We have explored some of the consequences of these notions of energy for the simplest
cases of a Schwarzschild and a de Sitter black hole. We considered also the different notions
of energy for the black hole itself. For that, the formalism that has proved useful is that
of isolated horizons. In that case there is a preferred notion of horizon energy that is
consistent with energy balance at infinity. If we use this notion of energy for the horizon,
then we get a fully consistent description of the energetics if we associate to the in-falling
object the notion of \emph{total} energy $\E$. In the de Sitter case, one has to appropriately modify
the notion of asymptotic infinity, but the qualitative picture remains intact. We have also
considered an alternative description where the energy associated to the particle is the one
given by an observer at rest. If we modify the notion of horizon energy to retain local energy
conservation, we are \new{led} to a very different mass for the horizon that is, however, not tailored
to the global notion of energy at infinity. \new{As we have shown, the two descriptions are nevertheless related by the Tolman redshift factor and assign the same change of area, and hence the same entropy, to the horizon; what is lost is the global energy balance.} One should note that, even when we have explicitly
considered only static vacuum solutions, one expects these results to be qualitatively similar
on other static black hole spacetimes, including matter, and for stationary solutions.

One could also wonder what would be the situation of a test matter field as defined by a
stress energy tensor $T_{ab}$. In this case, there is also a distinction between the special relativity
notion of energy momentum flux as seen by an observer $w^a$, given by $j^a=T^a{}_bw^b$, and the
\emph{conserved} current when a KVF $\xi^a$ is present, given by $\tilde j^a:=T^a{}_b\xi^b$. Note that since $\nabla_a\tilde j^a=0$,
the integral,
\begin{equation}
\E_{\rm matt}:=\int_\Sigma\dd^3x\,\sqrt{q}\,\tilde j^an_a=\int_\Sigma\dd^3x\,\sqrt{q}\,T_{ab}\,\xi^an^b \label{eq:Ematt}
\end{equation}
over a Cauchy surface $\Sigma$, is conserved. Again, the difference between the two integrands
depends on the redshift factor $V=|\xi_a\xi^a|^{1/2}<1$ that accounts for the gravitational (potential) energy. In this case, it affects the local notion of energy flux density and distinguishes it from the (non-conserved) generalization of the SR notion of energy defined by
$\new{E}_{\rm matt}:=\int_\Sigma\dd^3x\,\new{\sqrt{q}}\,T_{ab}\,n^an^b$. It should be noted that the conserved energy (\ref{eq:Ematt}) is closely
related to the quantity defined by Komar. If we consider the issue of matter flowing into
the black hole, then the relevant quantity representing flux across the horizon will be $T_{ab}\xi^a$,
and \emph{not} the quantity $T_{ab}w^a$. The relation between energy flux and the change in mass of the
black hole will again depend on the `observer' we choose to describe the process. Compatibility with the total notion of energy will select the quantity $\E$, while the `local' viewpoint
will select the quantity $E$ as in~\cite{gp}. This discussion also \new{points} to the challenges that the
proposal of~\cite{gp} faces in order to be physically compelling.

\section*{Acknowledgments}

This work was in part supported by DGAPA-UNAM IN103610 grant, by CONACyT 0177840
and 0232902 grants, by the PASPA-DGAPA program, by NSF PHY-1403943 and PHY-1205388 grants, and by the Eberly Research Funds of Penn State.

\end{document}